\documentstyle[epsfig,aps,prb,preprint,graphicx,amssymb,latexsym]{revtex}
\begin{document}
\draft

\title{Liquid-Liquid Phase Transition for an Attractive Isotropic Potential
with Wide Repulsive Range}

\author{Gianpietro Malescio$^1$, Giancarlo Franzese$^{2,3}$, Anna
Skibinsky$^3$, Sergey V. Buldyrev$^{3,4}$, and H. Eugene
Stanley$^3$}

\address{$^1$Dipartimento di Fisica, Universit\`a di Messina and
Istituto Nazionale Fisica della Materia, 98166 Messina, Italy\\
$^2$Departament de F\'{\i}sica Fonamental, Facultat de
F\'{\i}sica, Universitat de Barcelona, Diagonal 647, 08028 Barcelona, Spain\\
$^3$Center for Polymer Studies and Department of Physics, Boston
Univ., Boston, MA 02215, USA\\
$^4$Department of Physics, Yeshiva University, 500 W 185th St.,
New York, New York, 10040,USA\\ }

\date{10 november 2004}

\maketitle

\begin{abstract}

Recent experimental and theoretical results have shown the
existence of a liquid-liquid phase transition in isotropic
systems, such as biological solutions and colloids, whose
interaction can be represented via an effective potential with a
repulsive soft-core and an attractive part. We investigate how the
phase diagram of a schematic general isotropic system, interacting
via a soft-core squared attractive potential, changes by varying
the parameters of the potential. It has been shown that this
potential has a phase diagram with a liquid-liquid phase
transition in addition to the standard gas-liquid phase transition
and that, for a short-range soft-core, the phase diagram resulting
from molecular dynamics simulations can be interpreted through a
modified van der Waals equation. Here we consider the case of
soft-core ranges comparable with or larger than the hard-core
diameter. Because an analysis using molecular dynamics simulations
of such systems or potentials is too time-demanding, we adopt an
integral equation approach in the hypernetted-chain approximation.
Thus we can estimate how the temperature and density of both
critical points depend on the potential's parameters for large
soft-core ranges. The present results confirm and extend our
previous analysis, showing that this potential has two fluid-fluid
critical points that are well separated in temperature and in
density only if there is a balance between the attractive and
repulsive part of the potential. We find that for large soft-core
ranges our results satisfy a simple relation between the
potential's parameters.

\end{abstract}

\bigskip
\bigskip


\section{Introduction}

The phase diagram of a typical monatomic substance is comprised of
solid and fluid phases, with the fluid phase separating below the
critical point into gas and liquid phases. The prototype of such
substances are simple (i.e., argon-like) fluids. Interparticle
interactions in these systems can be appropriately described by
the well-known Lennard-Jones potential. Other simple models---such
as those described by the hard-sphere square-well potential or by
the hard-sphere-Yukawa potential---exhibit similar phase diagrams.
All these potentials consist in a short-range harshly repulsive
core plus a longer-ranged attraction. New insights into the
relationship between phase diagrams and interparticle interaction
emerged recently from the finding that when the range of the
attractive component is sufficiently small, the liquid phase and
the gas-liquid critical point become metastable with respect to
crystallization
\cite{LEKKERKERKER,BOLHUIS-SQW,BOLHUIS-SQW2,Frenkel,Stell4}.
Shouldered potentials, i.e., potentials with a hard core and a
finite repulsive shoulder, exhibit even more exotic phase
diagrams. Simulations and theories showed that such potentials may
give origin to non-trivial phase behaviors, such as isostructural
solid-solid transitions and liquid-liquid transitions
\cite{YOUNG,BOLHUIS-SC,STELL2,HEMMER-SC,franzese1,malescio,%
PhysA,malescio2,franzese2,anna}. The key to this complex phase
behavior resides in the peculiar penetrability of the repulsive
core, a feature that gives rise to a {\it density-dependent}
effective interaction.

The possible existence of a liquid-liquid phase transition for
single-component systems in particular has received considerable
attention in recent years. A direct evidence of this phenomenon
has been observed on the experimental side in liquid phosphorous
\cite{Katayama,Monaco}. Experimental data consistent with a
liquid-liquid phase transition have been presented for other
single-component systems such as water
\cite{mishima_2000,bellisent,soper}, silica \cite{Lacks,ANGELL},
carbon \cite{Thiel}, selenium \cite{Brazhkin98}, and cobalt
\cite{Vasin}, among others. A recent theory has shown that a phase
transition occurring in a solution of DNA-coated colloids can be
understood in terms of the liquid-liquid phase transition
\cite{Lukatsky}. A liquid-liquid critical point has also been
predicted by simulations for specific models of supercooled water
\cite{Poole,FMS,slt}, carbon \cite{G}, phosphorous
\cite{MORISHITA}, supercooled silica
\cite{ANGELL,Saika-Voivod,SHRI_PP}, and hydrogen \cite{hydrogen}.

We have recently shown through molecular dynamics (MD)
simulations\cite{franzese1} that a system of particles interacting
through an isotropic potential with an attractive well and a
repulsive component consisting of a hard core plus a finite
shoulder may possess a high-density liquid phase and a low-density
liquid phase. Potentials with such characteristics were used to
model interactions in a variety of systems including liquid
metals, metallic mixtures, electrolytes, and colloids, as well as
anomalous liquids, like water and silica
\cite{hs,Deb1,SLK,SA,L75,DRB91,SHG93,Deb2,ssbs,jagla}.

In spite of the simplicity of the model, the physical mechanism
that causes the liquid-liquid transition for a potential with a
hard core plus a repulsive shoulder and an attractive well is not
easy to assess since it arises from an interplay of the different
components of the interparticle interaction. To disentangle the
role of each component it is necessary to investigate the
dependence of the phase stability of the fluid on the potential
parameters. This task was undertaken in Ref.~\cite{anna}, where
the results of MD calculations performed for several sets of
parameters were presented. The resulting behavior of the critical
points was interpreted through a modified van der Waals equation
(MVDWE) \cite{anna}, a mean field approach assuming that the
effect of the repulsive shoulder at different densities $\rho$ and
temperatures $T$ can be taken into account by an effective
excluded volume depending on both $\rho$ and $T$. However, the
analysis was limited to cases where both the soft-core range and
the attractive range are smaller than the hard-core range $a$ and
the total interaction range does not exceed $2.6a$. Indeed, MD
becomes quite time-demanding for larger interaction ranges.
Nevertheless, there are cases such as biological solutions and
colloids where the soft-core range could be as large as the
hard-core or even larger \cite{ultrasoft}. For this reason and to
gain a better understanding of the role played by each component
of a soft-core's attractive potentials, we will now explore how
the phase diagram changes when the soft-core range exceeds the
hard-core diameter.

For this we use the hypernetted-chain (HNC) integral equation for
the radial distribution function \cite{HM}. This approach can be
considered in many respects as an intermediate between MD and
MVDWE. In fact, the HNC equation is far less time-consuming than a
simulation, but is by no means as accurate. On the other hand,
being a microscopic theory, it is not based on the assumption
typical among mean field approaches such as the MVDWE that
particles experience a uniform attractive potential. Hence it is
intrinsically more accurate. Since the HNC equation provides a
fast way of estimating the position of the critical points, we
perform an extensive investigation in the space of the potential
parameters, considering an extremely ample number of combinations.
Thus we are able to frame previous results into a wider
perspective and obtain a better understanding of the physical
mechanism leading to a liquid-liquid transition in one-component
fluids.

\begin{figure}
\centering
\includegraphics[width=10.0cm,angle=0]{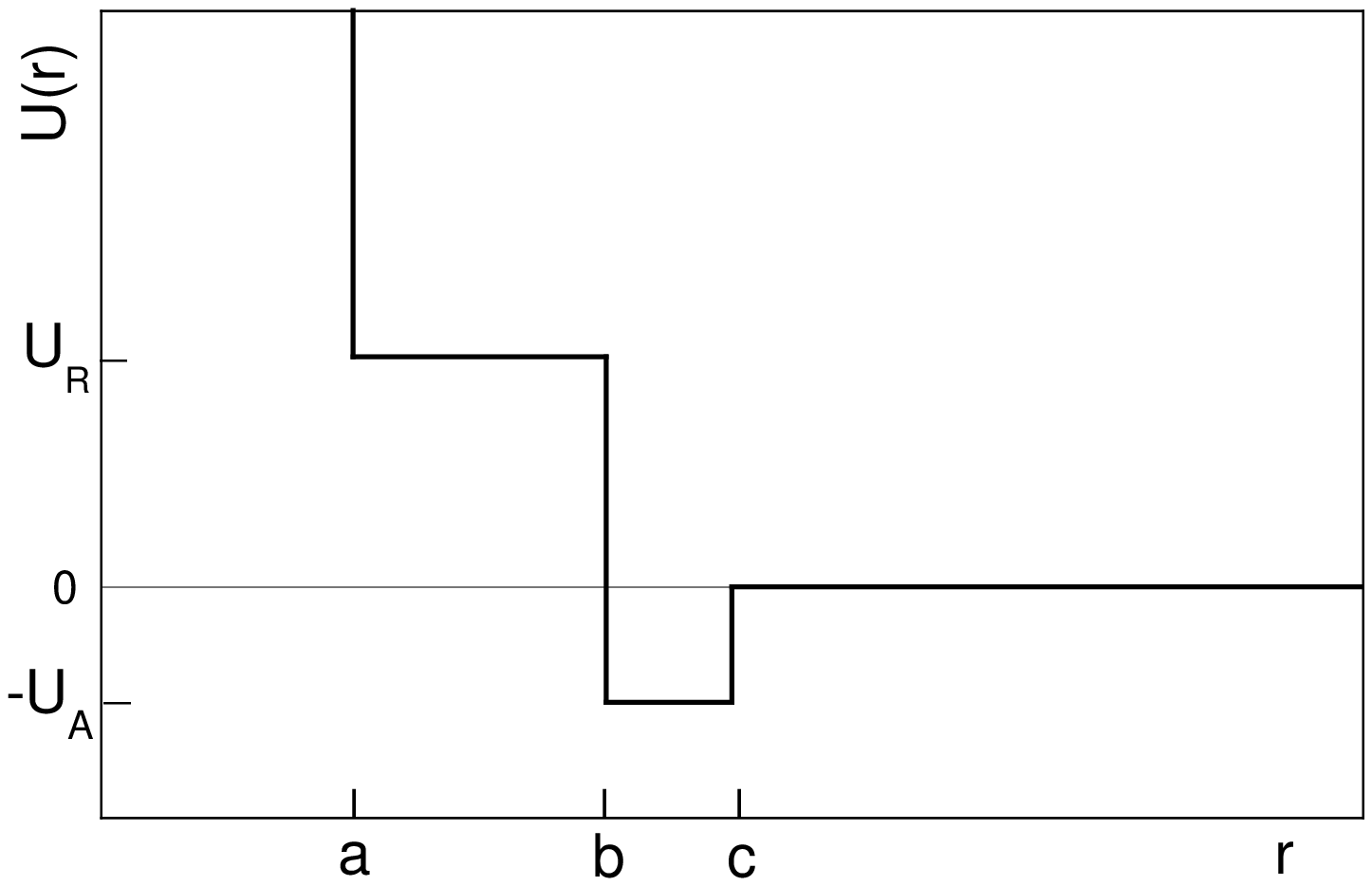}
\caption{General shape of the attractive soft-core potential used
in this work, with hard-core distance $a$, soft-core distance $b$,
interaction range $c$, attractive energy $U_A$, and repulsive
energy $U_R$. We use $U_R/U_A$, $w_A=c-b$, and $w_R=b-a$ as
independent parameters.} \label{1}
\end{figure}

Our results show that the high-density critical point can be found
only when there is a balance between the attractive part and the
repulsive part of the potential. In Ref.~\cite{anna} this balance
was expressed through the mean field {\it strength of attraction},
a parameter proportional to the attractive range $w_A/a$
(Fig.~\ref{1}) and inversely proportional to the repulsive energy
$U_R$, for fixed attractive energy $U_A$. Here we find an
approximated relation between $U_R/U_A$ and $w_A/w_R$ that is
well-verified for large soft-core ranges and quantifies the ideal
balance between the repulsive and the attractive components of the
potential more effectively. Our results show that the
liquid-liquid phase transition could be found in systems with
small repulsion if the attraction is small as well, with
$U_R/U_A\propto w_A/w_R$, and in systems with wide repulsion, with
$U_R/U_A\propto 3 w_A/w_R$. Typical systems with these
characteristics could be colloids, where the effective repulsion
and attraction can be regulated \cite{Lukatsky,Baksh}.

\section{The attractive soft-core potential}

A soft-core potential with an attractive interaction at large
distances was first proposed by Hemmer and Stell \cite{hs} to
understand the possibility of the solid-solid critical point in
materials such as Ce and Cs and was studied through an exact
analysis in 1D. Other soft-core potentials with an attractive well
were proposed and studied with approximate methods or with
numerical simulations in 2D to rationalize the properties of
liquid metals, alloys, electrolytes, colloids, and the
water anomalies \cite{Deb1,%
SLK,SA,L75,DRB91,SHG93,Deb2,ssbs,franzese1,malescio,jagla,%
PhysA,franzese2}.

The peculiarity of such potentials is the presence of two
repulsive length scales. This feature is typical of systems with
core-corona architecture such as, for example, star polymers.
However, isotropic soft-core potentials have also been proposed as
effective potentials resulting from an average over the angular
degrees of freedom for systems where the distance of the minimum
approach between particles depends on their relative orientation.
Thus, in some respects, they have been considered
\cite{SHG93,ssbs,jagla} as simplified models of complex
anisotropic interactions, such as those resulting from the
hydrogen bonding between water molecules.

The model potential considered in this paper is similar to that
investigated in Refs.\cite{franzese1,malescio,franzese2,anna} and
is an isotropic pair potential with two characteristic short-range
repulsive distances: one associated with the hard-core exclusion
between two particles and the second with a weak repulsion
(soft-core), which can be overcome at large pressure. More
precisely, our pair potential $U(r)$ (Fig.~\ref{1}) consists of a
hard-core of radius $a$, a repulsive square shoulder of height
$U_R$ extending from $r=a$ to $r=b$, and an attractive component
having the form of a square well of energy $-U_A<0$ extending from
$r=b$ to $r=c$ (here $r$ is the interparticle distance). Choosing
$a$ and $U_A$ as length and energy units respectively, this
potential depends on three free parameters: the width of the
soft-core $w_R/a \equiv (b-a)/a$, the width of the attractive well
$w_A/a \equiv (c-b)/a$, and the soft-core energy $U_R/U_A$.

Our aim is to understand how the position of the critical points
in the thermodynamic plane changes upon varying the parameter
values. In Ref.~\cite{anna} we investigated a number of cases with
$w_R<a$ and $w_A<a$ through MD simulations and presented a mean
field approach with an MVDWE to rationalize the results. However,
the MD analysis of potentials with $w_R\geq a$ would require very
large computation times, therefore we study this case with
integral equations in the HNC approximation, which represents a
compromise between accuracy and economy.

\section{The hypernetted chain integral equation approach}

The spatial distribution of a system of particles may be
conveniently described by the radial distribution function
$g(r)$\cite{HM}, a quantity directly measurable by scattering
experiments and related to the thermodynamic properties of the
fluid. One of the theoretical approaches most used to calculate
this function is represented by integral equations. These are
based on the so-called Ornstein-Zernike (OZ) relation which
relates the total pair correlation function $h(r)\equiv g(r)-1$ to
the direct correlation function $c(r)$, which describes the
contribution coming from the direct interaction between two
particles at distance $r$. Both $h(r)$ and $c(r)$ are unknown
functions, thus to solve the OZ relation, one needs another
relation ({\it closure}) between these two functions. This is by
necessity an approximate relation and it is obtained in the HNC
equation\cite{HNC1} by neglecting the sum over a specific class of
diagrams in the diagrammatic expansion of $g(r)$\cite{HM}. In
principle, this approximation is expected to work better at lower
$\rho$, where the direct correlation function $c(r)$ is more
relevant than the correlation propagated through the other
particles.

The solution of the system formed by the OZ relation plus the HNC
closure is obtained through a numerical iterative procedure which
is stopped when the difference between two consecutive elements of
the succession generated (in the space of the distribution
functions) is smaller than some given value. Based on the standard
iterative procedure, different algorithms can be used to improve
the accuracy and rapidity of convergence of the HNC equation's
numerical solution. However, independent of the algorithm used,
there is a region in the $\rho$-$T$ plane where no solution can be
found, i.e., for any $\rho$, there is a $T$ below which the
numerical algorithm does not converge. This defines an instability
line (IL) of the theory in the $\rho$-$T$ plane. The nature of the
locus of instabilities of the HNC equation and its relationship
with the spinodal line of the fluid was the object of extensive
investigation\cite{HNC2} which showed that though the isothermal
compressibility increases considerably when the instability line
is approached from above, there is no real divergence of this
quantity. Thus, identifying the IL of the HNC equation with the
spinodal line of the fluid, which is characterized by a diverging
compressibility, is not possible. Another well-known deficiency of
the HNC is its thermodynamical inconsistency. This can at least be
partially removed through suitable modifications of the equation
which, however, make the method considerably less rapid.

In spite of the above limitations, knowledge of the instability
region of the HNC equation may allow us to estimate the topology
of the region of spinodal decomposition of the fluid. In
particular, for the potential defined in Sect.~II, with parameters
$w_R/a=1$, $w_A/a=0.2$, and $U_R/U_A=0.5$, it was found that the
IL is qualitatively similar to the spinodal line calculated
through MD calculations\cite{franzese2}. More precisely, the
density and temperature of the low-density critical point
estimated through the HNC equation are in satisfactory agreement
with simulation results, while the density of the second-critical
point is overestimated by the theory\cite{franzese2}. This is not
surprising since the theory is an approximate one and becomes
progressively less accurate as the density increases. However, the
ability of the HNC equation to give account of the presence of two
critical points is, within the well-known limitations of the
theory, quite remarkable since the potential considered creates a
phase diagram that is definitely unusual for simple fluids. Thus,
studying the modifications of the IL as the potential parameters
are varied can yield approximate yet useful information on the
phase behavior of the fluid. In our calculations, functions are
evaluated on a grid with $M=2048$ discrete points $r_m=m\delta r$,
with $m=1,\dots, M$ and $\delta r/a=0.01$.

\section {Instability lines for the large repulsive range}

To disentangle the role of each component of the interparticle
interaction, the parameters of the potential are varied in our
investigation one at a time. First we keep the width $w_R$ of the
repulsive shoulder and the width $w_A$ of the attractive well
fixed and study the behavior of the IL, letting the height $U_R$
of the repulsive shoulder vary. The considered values of $U_R$
range from $-U_A$ to $\infty$. When $U_R/U_A=-1$, the potential
consists of a hard core of radius $a$ and a square well of width
$c-a$ (henceforth called {\it potential A}) whereas, when $U_R
\rightarrow \infty$, the potential has a hard core of radius $b$
and a square well of width $c-b$ ({\it potential B}). When the
shoulder height increases, the potential gradually changes from
potential A to potential B starting from $U_R/U_A=-1$. In any
intermediate configuration, the potential has a penetrable finite
repulsive shoulder. It is also possible to go from potential A to
B following a different ``path'', i.e., by increasing the hard
core radius from $a$ to $b$. In this case, however, the potential
always consists of an impenetrable hard-core (of varying radius)
plus an attractive square well.

\begin{figure}
\centering
\includegraphics[width=16.0cm,angle=0]{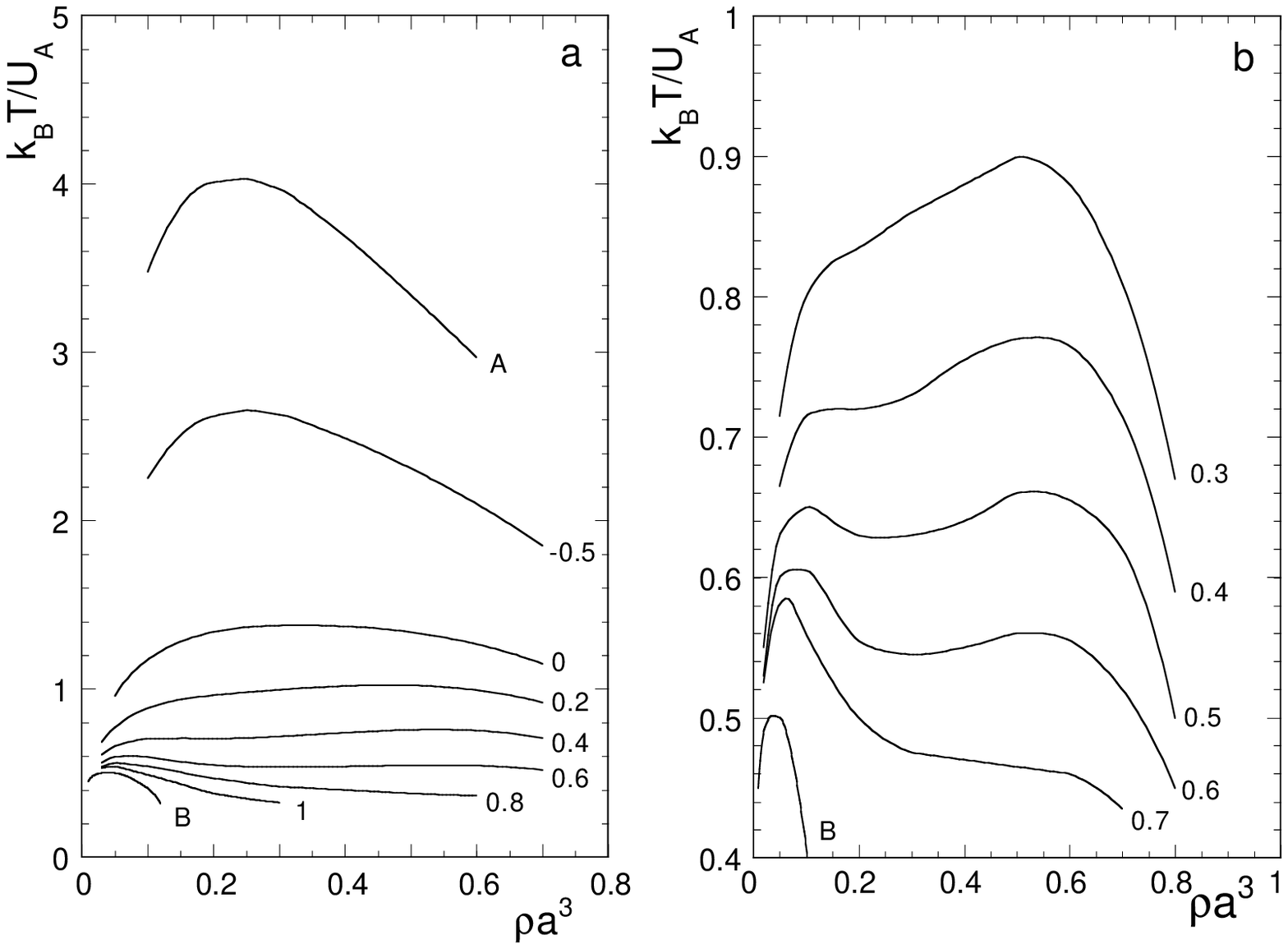}
\caption{Instability line of the HNC equation for the potential in
Fig.~\ref{1} with parameters $w_R/a=1$, $w_A/a=0.2$, and for (from
top to bottom in panel a) $U_R/U_A=-1$, -0.5, 0, 0.2, 0.4, 0.6,
0.8, 1, $\infty$, and (from top to bottom in panel b)
$U_R/U_A=0.3$, 0.4, 0.5, 0.6, 0.7, $\infty$. Labels $A$ and $B$
marks the curves corresponding to potentials $A$ and $B$.}
\label{2}
\end{figure}

The IL is shown in Fig.~\ref{2} at fixed shoulder and well widths
($w_R/a=1$, $w_A/a=0.2$) for several values of the shoulder height
$U_R$. In the two limit cases corresponding to potentials A and B,
the IL exhibits a single maximum corresponding to a phase diagram
with a single liquid-gas critical point, a well-known behavior for
a fluid of hard spheres with an attractive well. The position of
the critical points in the $\rho,T$ plane is considerably
different in the two cases. The critical point corresponding to
potential B is at a lower temperature than that corresponding to
potential A, due to the weaker attraction, i.e., shorter
attractive range $c-b$ of potential B with respect to the largest
attractive range $c-a$ of potential A. Furthermore, the critical
density for potential B is smaller than that for potential A and
rescales as the hard-core volume $(a/b)^3$ of the two potentials.
We observe that this rescaling overshadows the shift of the
critical point toward higher densities due to the decrease of the
attraction range (e.g., see Appendix A in Ref.~\cite{anna}),
unless $b/a\simeq 1$.

As $U_R$ increases, starting from $U_R/U_A=-1$ (potential A), the
IL moves toward lower temperatures as a consequence of the overall
reduction of the interparticle potential's attractive component.
At the same time, the IL undergoes a topological change which
eventually yields a line with two maxima (Fig.~\ref{2}b). This
peculiar topology of the IL becomes most evident for intermediate
values of $U_R$ ($0.4 \leq U_R/U_A \leq 0.6$). As $U_R$ increases
further, the second maximum disappears and again the shape of the
IL becomes more and more similar to the shape typical of the
hard-core square-well potential. Thus, when $w_R$ and $w_A$ are
fixed, two maxima are observed in the IL only for a finite range
$U^{max}_R \leq U_R \leq U^{min}_R$.

In this range of values, as $U_R$ increases, the density $\rho_1$
of the low-density maximum becomes smaller, while that of the
other maximum $\rho_2$ slightly increases. The critical
temperatures $T_1$ and $T_2$ respectively corresponding to these
two maxima, decrease---this behavior being more evident for the
second maximum. These results agree with the behavior found with
MD simulations for the two critical points (reported in
Fig.~9g,~9h of Ref.~\cite{anna}). Thus for increasing $U_R$, the
two maxima move away from each other both in density and
temperature.

\begin{figure}
\centering

\includegraphics[width=16.0cm,angle=0]{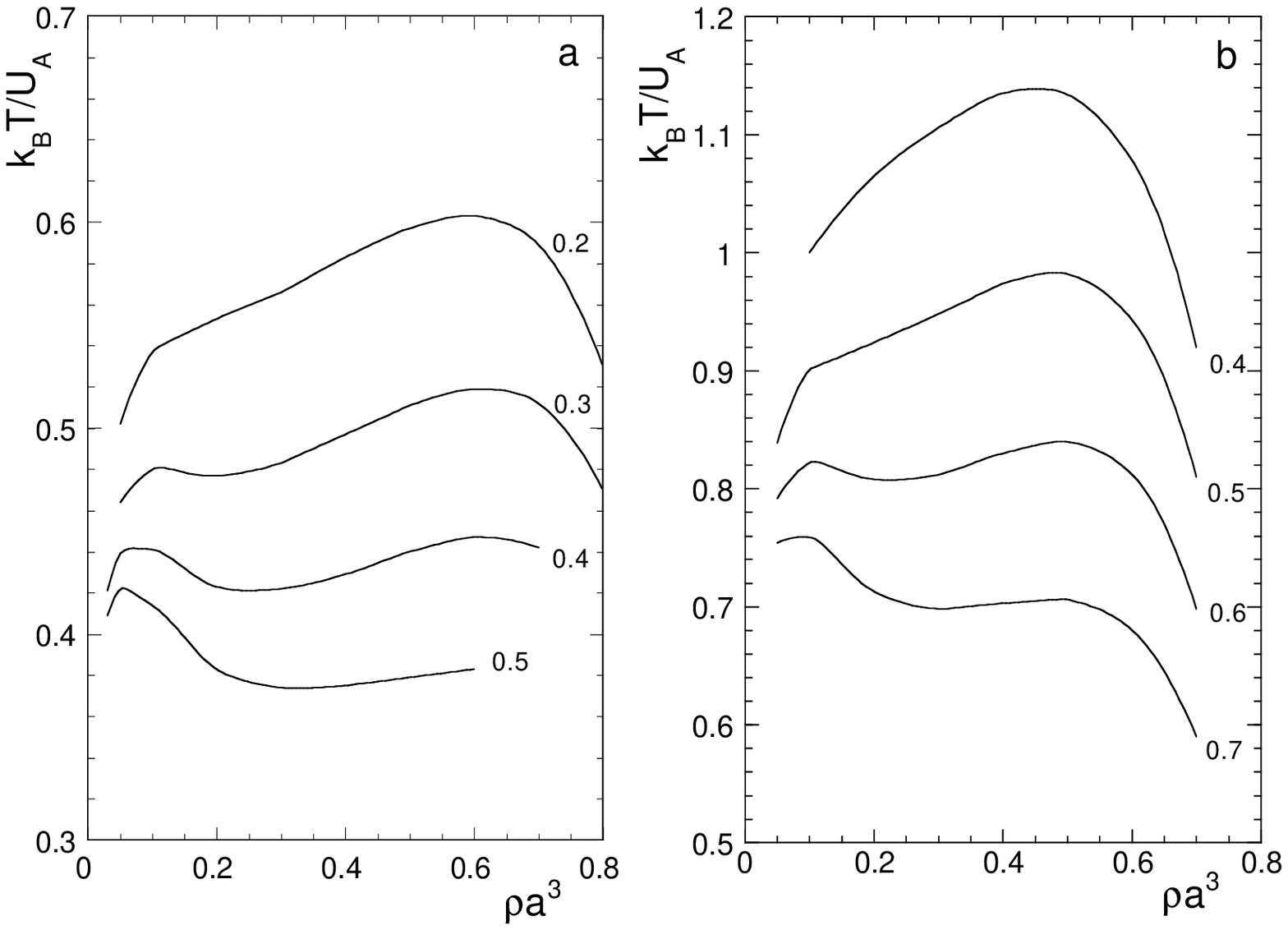}
\caption{Instability line of the HNC equation for the potential in
Fig.~\ref{1}: the parameters of the potential are, in panel a,
$w_R/a=1$, $w_A/a=0.1$, and (from top to bottom) $U_R/U_A=0.2$,
0.3, 0.4, 0.5; in panel b, $w_R/a=1$, $w_A/a=0.3$, and (from top
to bottom) $U_R/U_A=0.4$, 0.5, 0.6, 0.7.} \label{3}
\end{figure}

\begin{figure}
\centering
\includegraphics[width=16.0cm,angle=0]{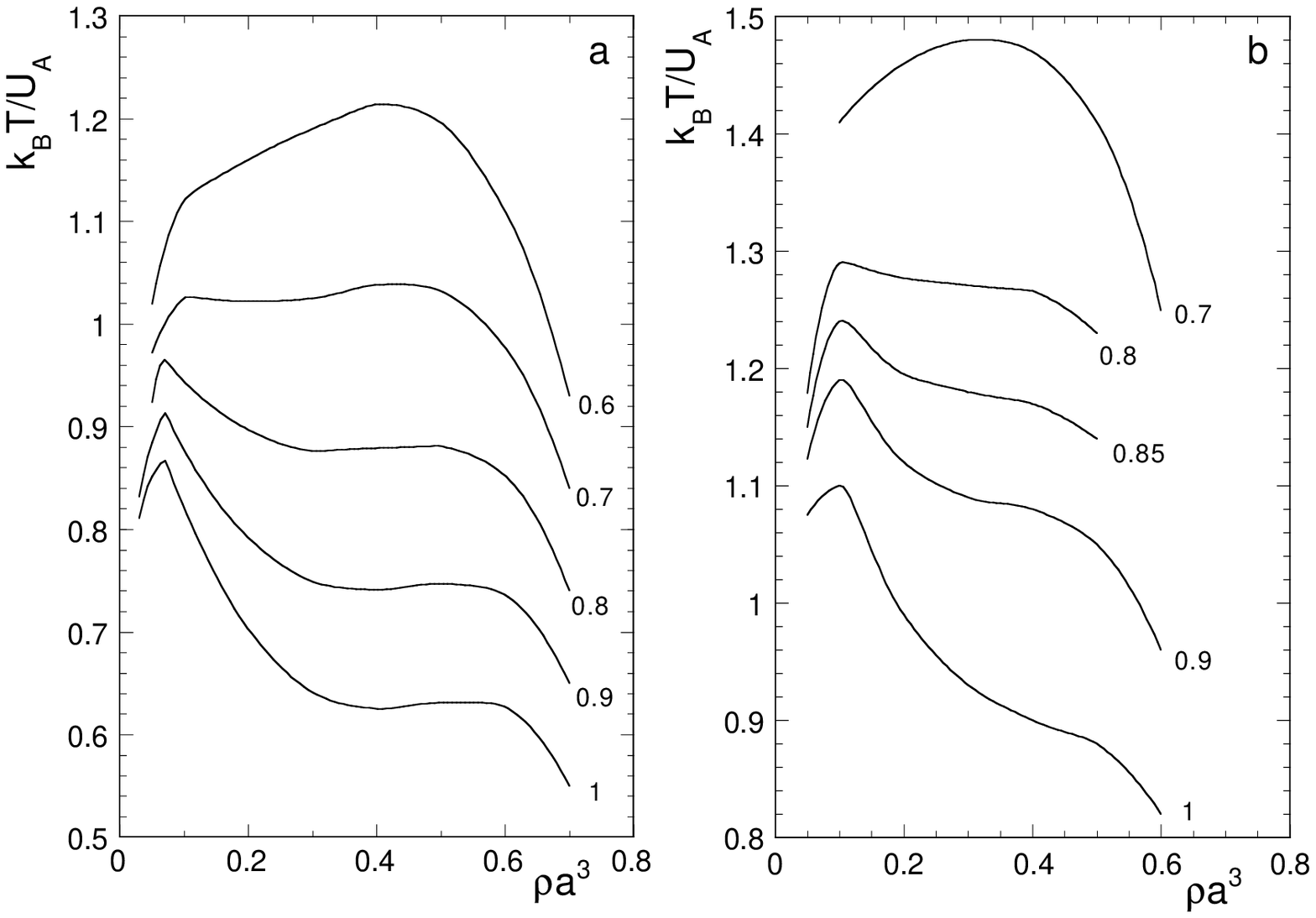}
\caption{Instability line of the HNC equation for the potential in
Fig.~\ref{1}: the parameters of the potential are, in panel a,
$w_R/a=1$, $w_A/a=0.4$, and (from top to bottom) $U_R/U_A=0.6$,
0.7, 0.8, 0.9, 1; in panel b, $w_R/a=1$, $w_A/a=0.5$, and (from
top to bottom) $U_R/U_A=0.7$, 0.8, 0.85, 0.9, 1.} \label{4}
\end{figure}

\begin{figure}
\centering
\includegraphics[width=16.0cm,angle=0]{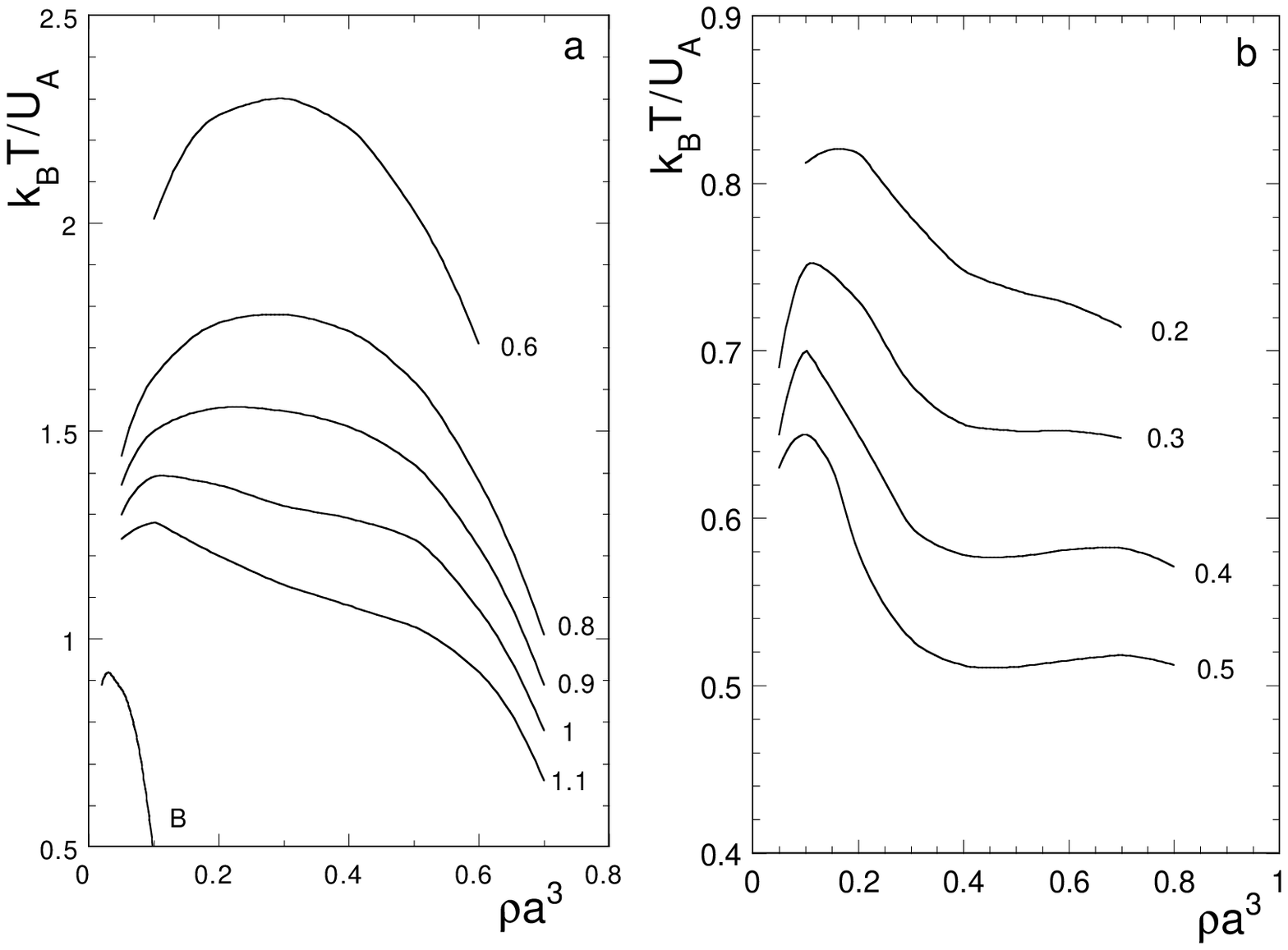}
\caption{Instability line of the HNC equation for the potential in
Fig.~\ref{1}: the parameters of the potential are, in panel a,
$w_R/a=1$, $w_A/a=0.6$, and (from top to bottom) $U_R/U_A=0.6$,
0.8, 0.9, 1, 1.1, $\infty$; in panel b, $w_R/a=0.8$, $w_A/a=0.2$,
and (from top to bottom) $U_R/U_A=0.2$, 0.3, 0.4, 0.5.} \label{5}
\end{figure}

We now consider a fixed shoulder width ($w_R/a=1$) and several
values of the well width ($w_A/a=0.1$, 0.2, 0.3, 0.4, 0.5, 0.6).
We calculate the IL for each of them, letting the height $U_R$ of
the shoulder vary (Figs.~\ref{3}a, \ref{2}, \ref{3}b, \ref{4}a,
\ref{4}b and \ref{5}a). The values of the shoulder height $U_R$
for which two maxima are observed increase with $w_A$. By
increasing $w_A$, the second maximum is less and less evident and
for large values of $w_A/a$ (Figs.~\ref{4}b and \ref{5}a) the
second maximum is not observed for any $U_R$. For small values of
$w_A$, the decrease of the attraction flattens the curve and the
second maximum becomes difficult to observe (Fig.~\ref{3}a).

For IL's with two maxima and the same $w_R/a=1$ and $U_R$ but
different $w_A$, both maxima move toward higher temperatures for
increasing $w_A$, due to the increased attraction. Moreover, by
increasing $w_A$, $\rho_2$ becomes smaller while $\rho_1$ does not
vary significantly. This behavior agrees with that predicted by MD
simulations for the two critical points (shown in Fig.~9a,9b of
Ref.~\cite{anna}).

In Fig.~\ref{5}b we show the behavior of the IL for $w_R/a=0.8$
and $w_A/a=0.2$. Comparing these results with those shown in
Fig.~\ref{2}, we observe that for fixed $w_A$ and $U_R$, as $w_R$
increases, $\rho_1$ and $T_1$ are almost constant and at variance
with MD results, but $\rho_2$ decreases and $T_2$ increases in
agreement with the MD simulations (see Fig.~9d,9e of
Ref.~\cite{anna}).

\begin{figure}
\centering
\includegraphics[width=16.0cm,angle=0]{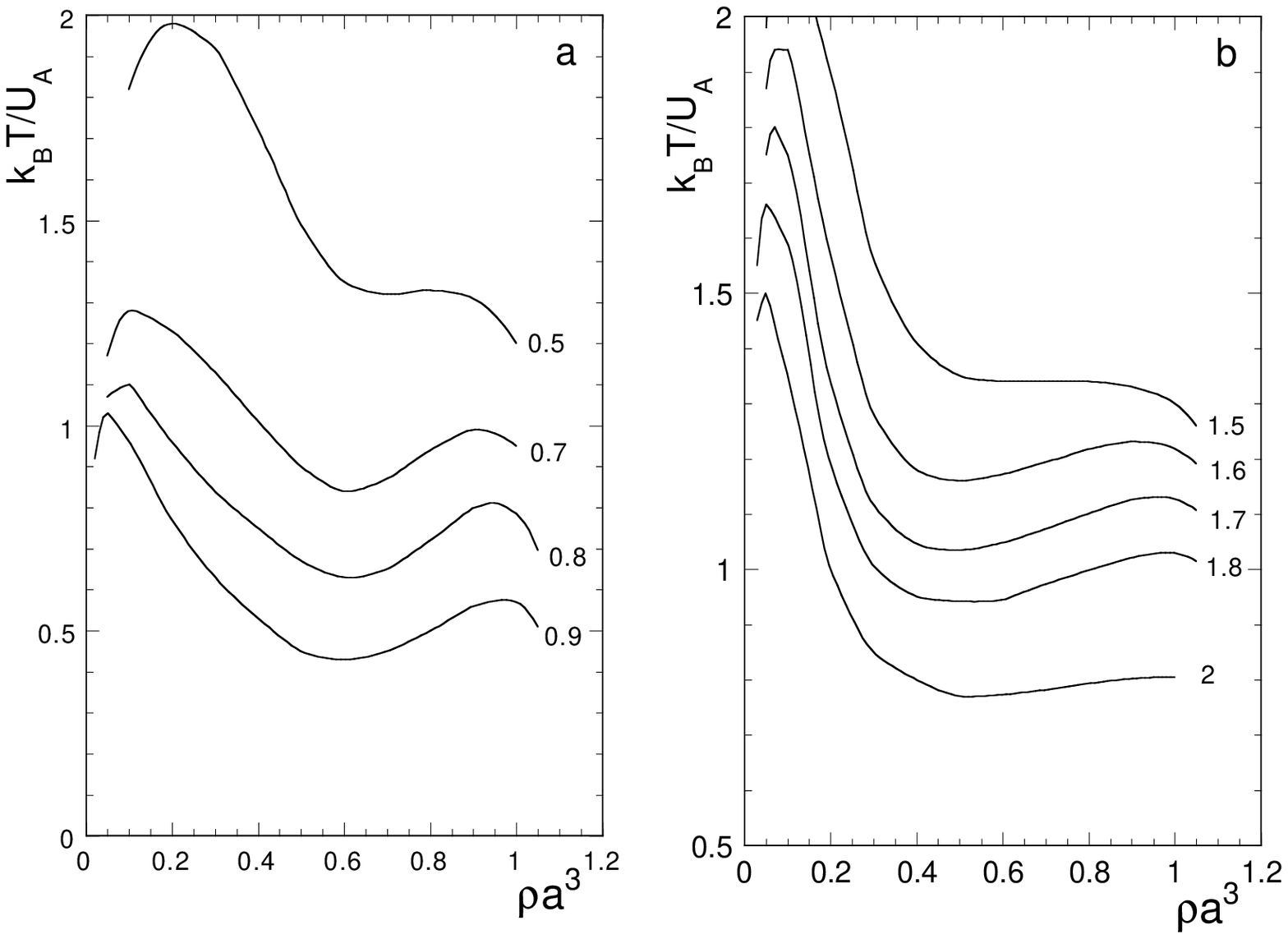}
\caption{Instability line of the HNC equation for the potential in
Fig.~\ref{1}: the parameters of the potential are, in panel a,
$w_R/a=1.5$, $w_A/a=0.5$, and (from top to bottom) $U_R/U_A=0.5$,
0.7, 0.8, 0.9; in panel b, $w_R/a=1.5$, $w_A/a=1$, and (from top
to bottom) $U_R/U_A=1.5$, 1.6, 1.7, 1.8, 2.} \label{6}
\end{figure}

\begin{figure}
\centering

\includegraphics[width=16.0cm,angle=0]{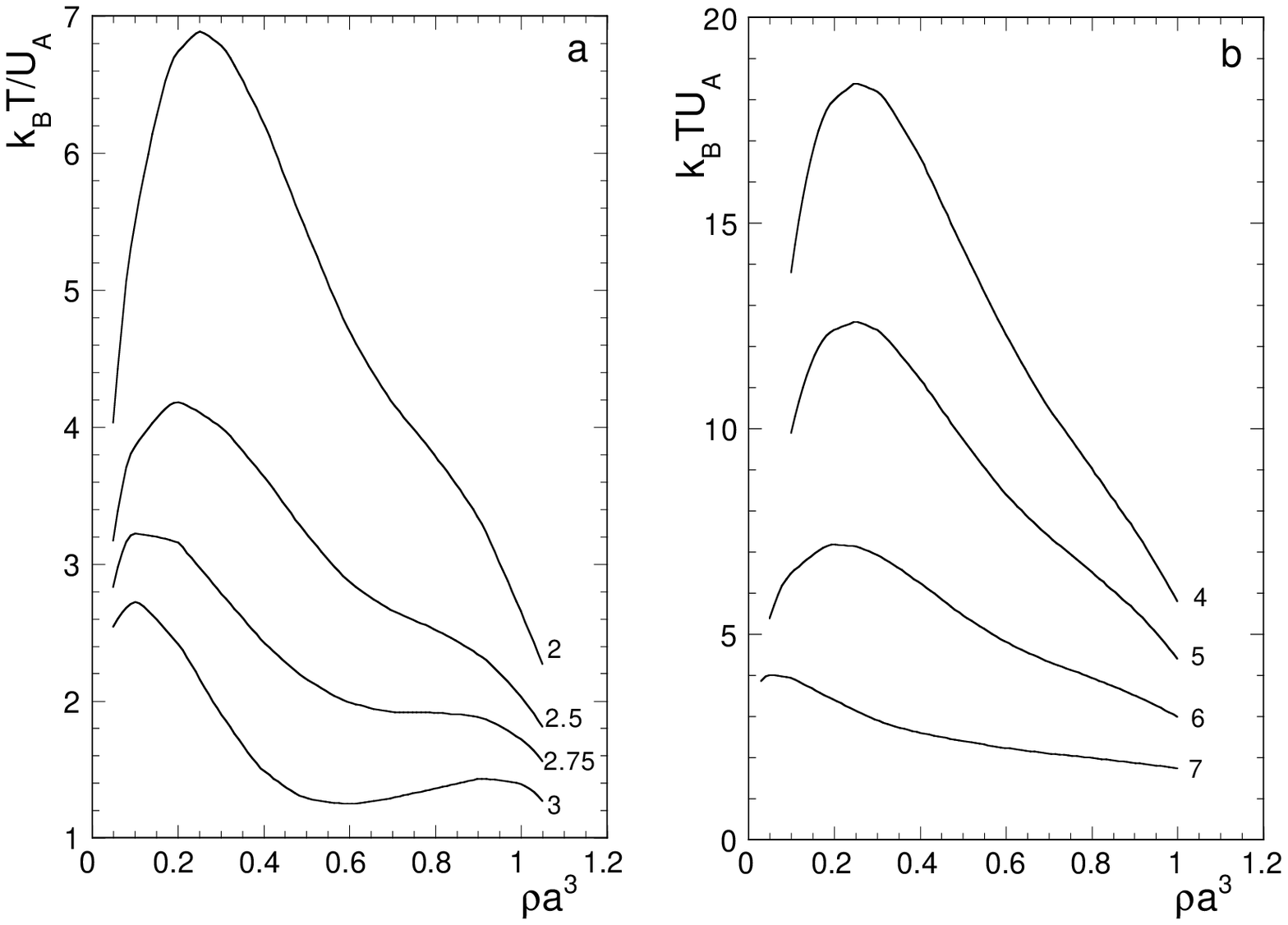}
\caption{Instability line of the HNC equation for the potential in
Fig.~\ref{1}: the parameters of the potential are, in panel a,
$w_R/a=1.5$, $w_A/a=1.5$, and (from top to bottom) $U_R/U_A=2$,
2.5, 2.75, 3; in panel b, $w_R/a=1.5$, $w_A/a=2.5$, and (from top
to bottom) $U_R/U_A=4$, 5, 6, 7.} \label{7}
\end{figure}

We next consider a potential with a wider repulsive shoulder
($w_R/a=1.5$) and several values of the well width ($w_A/a=0.5$,
1.0, 1.5, 2.5). The behavior of the IL (Figs.~\ref{6} and \ref{7})
for varying $U_R$ is quite similar to that observed in the
previous cases. For fixed $w_R$ and $w_A$, the IL shows only two
maxima in a finite range of values of $U_R$; these values increase
with $w_A$ and, for large values of $w_A$, the two-maxima topology
is not observed regardless of the value of $U_R$ (Fig.~\ref{7}b).
The range of values of $w_A$ in which we observe two maxima is
larger with respect to the case $w_R/a=1$.

For one particular set of parameters ($w_R/a=1.5$, $w_A/a=0.5$ and
$U_R/U_A=0.8$), it is possible to compare the results obtained
using the HNC equation with the phase diagram calculated through a
theoretical approach based on a thermodynamically consistent
integral equation \cite{malescio}. Once again it appears evident
that the main flaw of the HNC equation is to overestimate the
critical density of the second critical point.

\begin{figure}
\centering
\includegraphics[width=16.0cm,angle=0]{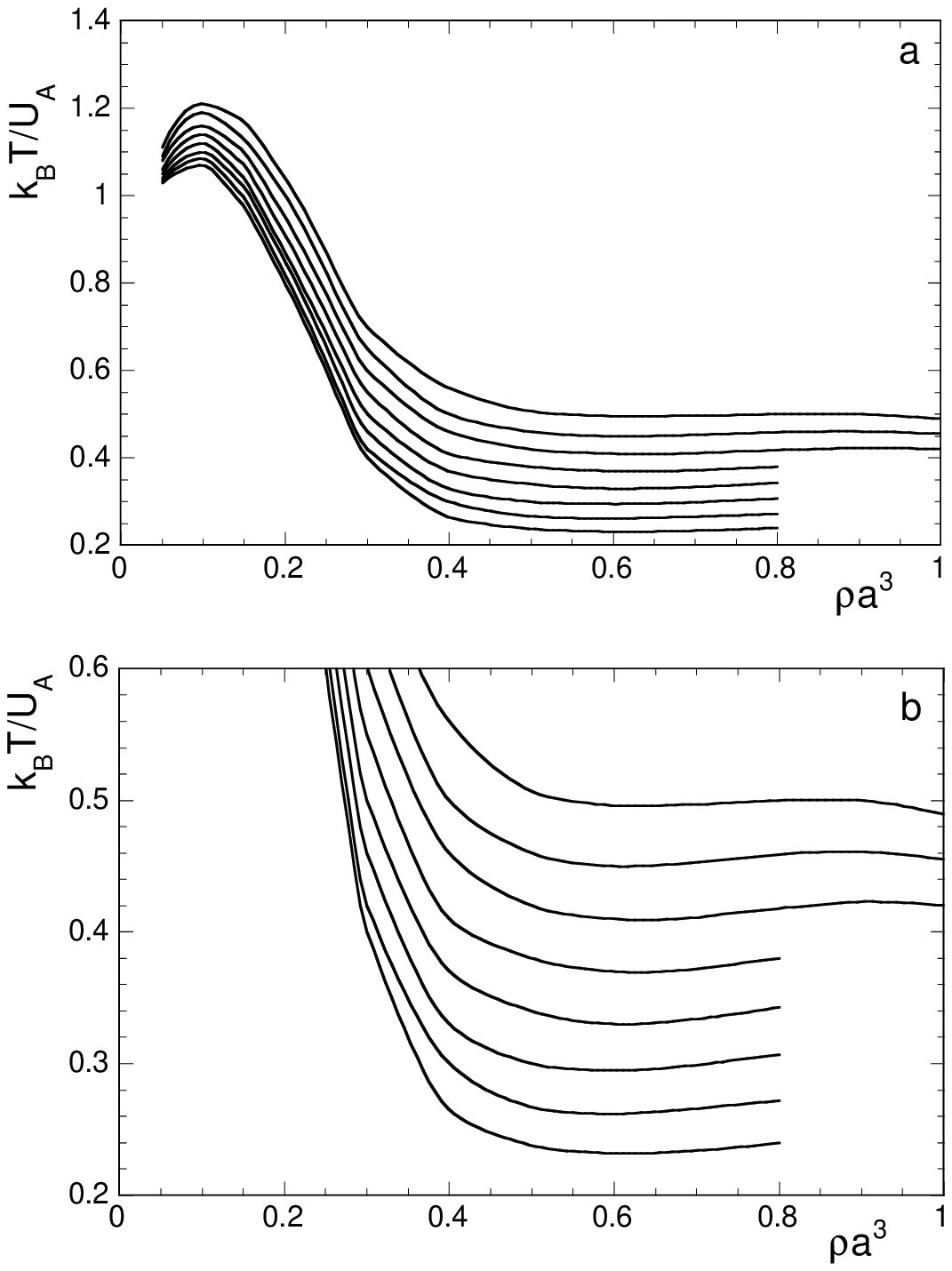}
\caption{Panel a: Instability line of the HNC equation for the
potential in Fig.~\ref{1} for $w_R/a=0.5$, $w_A/a=0.5$, and (from
top to bottom) $U_R/U_A=1$, 1.1, 1.2, 1.3, 1.4, 1.5, 1.6, 1.7;
panel b: an enlarged view at low $T$.} \label{8}
\end{figure}

However, a direct comparison of HNC results with those obtained
through MD simulations can be disappointing. For some of the
parameter sets investigated in Fig.~9 of Ref.~\cite{anna} the IL
shows only one maximum, while for others the two-maxima topology
is barely observable. As an example, we show the IL corresponding
to the parameters $w_R/a=0.5$, $w_A/a=0.5$ with $1.0\leq U_R/U_A
\leq 1.7$ (Fig.~\ref{8}). It was not possible to directly analyze
the value $U_R/U_A=2$ (considered in Ref.~\cite{anna}) since, in
this case, the HNC cannot be solved at high densities before any
considerable increase of the compressibility can be observed (in
general, this occurs when the finite repulsion is considerably
stronger than the attraction). The results obtained at slightly
smaller values of $U_R$ show, however, a non-monotonic behavior of
the IL, suggesting the presence of a liquid-liquid critical point.

\section {Discussion}

\begin{figure}
\centering
\includegraphics[width=11.0cm,angle=270]{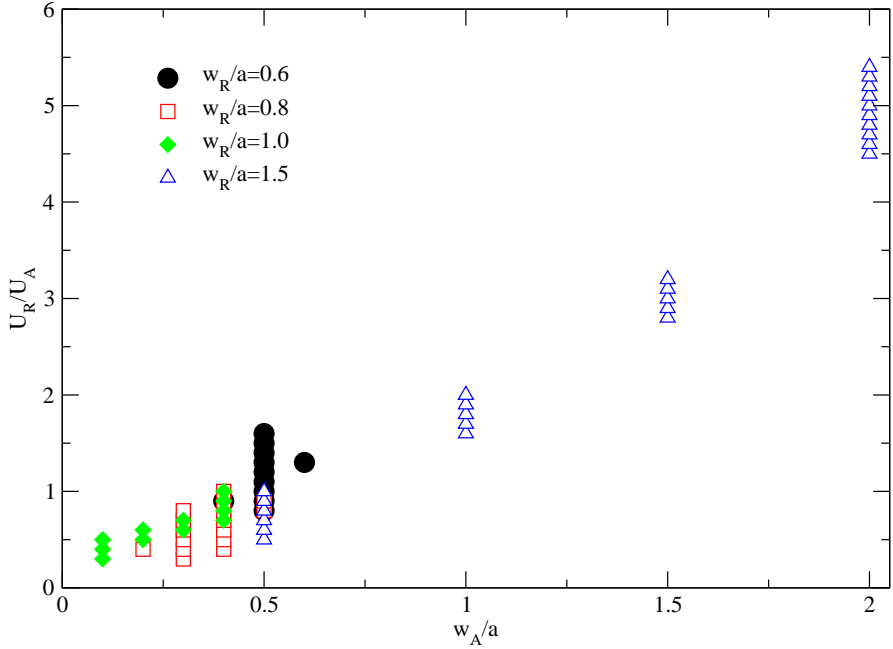}
\caption{Symbols mark the combinations of the potential's
parameters where the instability line, calculated by the HNC
approach, shows two maxima, suggesting the presence of two
fluid-fluid critical points. Sets with $w_R/a=0.6$ (circles) were
investigated for $0\leq U_R/U_A\leq 2$ and $0 \leq w_A/a \leq 1$;
sets with $w_R/a=0.8$ (squares) and with $w_R/a=1.0$ (diamonds)
for $0\leq U_R/U_A\leq 1.2$ and $0 \leq w_A/a \leq 0.6$; sets with
$w_R/a=1.5$ (triangles) for $0\leq U_R/U_A\leq 6$ and $0 \leq
w_A/a \leq 3$. Parameters outside these regions have not been
investigated.} \label{9}
\end{figure}

The overall behavior of the IL's is synthesized in Fig.~\ref{9}
which shows, for different values of the shoulder width
($w_R/a=0.6$, 0.8, 1.0, 1.5), the points in the $w_A$, $U_R$ plane
where two maxima are found by using the HNC approximation. We
observe that both ranges of $w_A$ and $U_R$ where two maxima are
observed increase with $w_R$.

\begin{figure}
\centering
\includegraphics[width=11.0cm,angle=270]{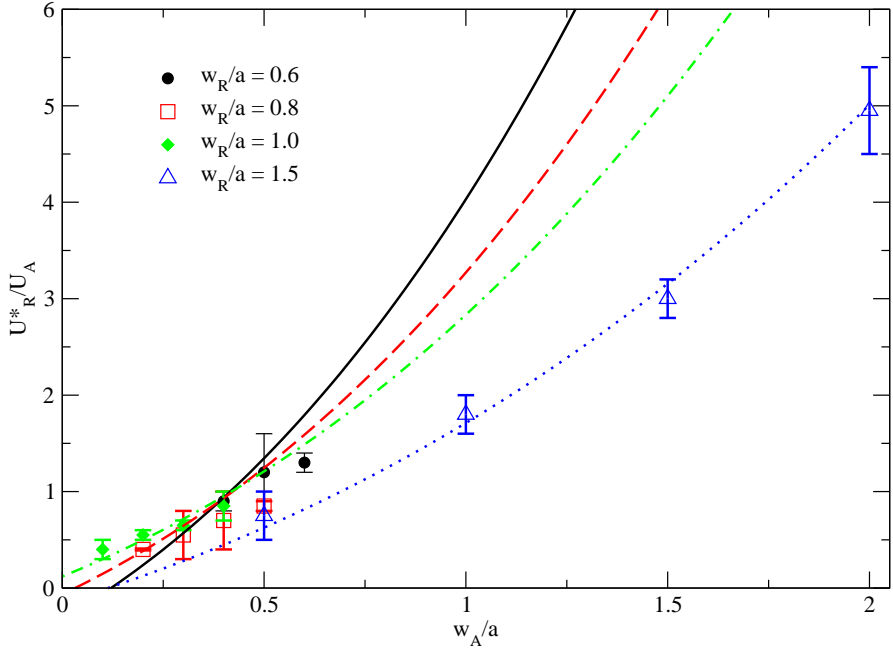}
\caption{Symbols mark the middle points $U^*_R$ of intervals of
$U_R$ in Fig.~\ref{9} for $w_R/a=0.6$ (set 1 denoted by circles),
$w_R/a=0.8$ (set 2 denoted by squares), $w_R/a=1.0$ (set 3 denoted
by diamonds), $w_R/a=1.5$ (set 4 denoted by triangles). Error bars
represent the interval in Fig.~\ref{9}. Lines are one-parameter
fits with Eq.~(\ref{mf}) of the sets: for set 2 (squares) the
fitting parameter is $A/(U_AV_{SC})=0.31$ (dashed line); for set 3
(diamonds) the fitting parameter is $A/(U_AV_{SC})=-0.84$
(dot-dashed line); for set 4 (triangles) the fitting parameter is
$A/(U_AV_{SC})=2.23$ (dotted line). Since we only have three
points for set 1, to avoid a fit with a large indeterminacy on the
parameters we arbitrarily chose $A/(U_AV_{SC})=1$ to show that the
data are consistent with Eq.~(\ref{mf}).} \label{10}
\end{figure}

The general behavior of $U_R$ as a function of $w_A$ at constant
$w_R$ can be rationalized by using the modified van der Waals
approach (MVDWE) presented in Ref.~\cite{anna}. First we
approximate the interval of values of $U^{min}_R \leq U_R \leq
U^{max}_R$ for each $w_R$ and $w_A$ in Fig.~\ref{9} with its
middle point $U_R^*=(U^{max}_R+U^{min}_R)/2$ (Fig.~\ref{10}).
Next, we recall from Ref.~\cite{anna} the relation between the
potential's parameters and the {\it strength of attraction} $A$, a
parameter increasing with $w_A/a$ and decreasing with $U_R/U_A$.
In particular, as $T\rightarrow \infty$ it is
\begin{equation}
A=U_Av_A-U_Rv_R, \label{eq1}
\end{equation}
with
\begin{equation}
v_A=\frac{2\pi}{3}[(a+w_R+w_A)^3-(a+w_R)^3], \label{va}
\end{equation}
and
\begin{equation}
v_R=\frac{2\pi}{3}[(a+w_R)^3-a^3]. \label{vr}
\end{equation}
The relation $U_R/U_A=v_A/v_R-A/(U_Av_R)$ can be rewritten as
\begin{equation}
\frac{U_R}{U_A}=\frac{V_{SC}}{V_{SC}-V_{HC}}\left[ -\frac{A}{U_A
V_{SC}} +3\frac{R_{HC}}{R_{SC}}\frac{w_A}{a}
+3\frac{S_{HC}}{S_{SC}}\left(\frac{w_A}{a}\right)^2
+\frac{V_{HC}}{V_{SC}}\left(\frac{w_A}{a}\right)^3 \right],
\label{mf}
\end{equation}
where the coefficients
\begin{eqnarray}
V_{HC} = \frac{2\pi}{3}a^3 ~, V_{SC} = \frac{2\pi}{3}(a+w_R)^3 ~,
\frac{S_{HC}}{S_{SC}} = \frac{a^2}{(a+w_R)^2} ~,
\frac{R_{HC}}{R_{SC}} = \frac{a}{a+w_R} \label{coefficients}
\end{eqnarray}
are the volumes, and the ratios of the surfaces and radii of the
hard core (HC) and the soft core (SC), respectively, and all
depend only on the parameter $w_R/a$. Hence, at a fixed value of
$w_R$, the function $U_R(w_A)$ in Eq.~(\ref{mf}) only has $A$ as
an unknown parameter.

This MVDWE prediction can be verified by using the HNC results. In
Fig.~\ref{10}, Eq.~(\ref{mf}) is used to fit the values of
$U_R^*(w_A)$ resulting from HNC calculations for different values
of $w_R$, with $A$ as the only fitting parameter. As expected from
Eq.~(\ref{mf}), when $w_A/a<1$, the leading order in $U_R^*(w_A)$
is linear, while when $w_A/a>1$ (corresponding to larger $w_R$),
the non-linear behavior is evident. Fig.~\ref{10} also shows that,
by increasing $w_R$, the coefficients of the third-degree
polynomial in $w_A$ decrease as predicted by Eqs.~(\ref{mf}) and
(\ref{coefficients}).

Moreover, the fitting parameter $A$ in Fig.~\ref{10} shows a
non-monotonic behavior with $w_R$. This is consistent with the
MVDWE prediction in Ref.~\cite{anna} that $\partial A/\partial
w_R$ may have different signs, depending on the other parameters.
Therefore, Eqs.~(\ref{mf}) and (\ref{coefficients}) give us a fair
description of how the three parameters $U_R$, $w_R$, and $w_A$
are related to each other when the phase diagram has two critical
points at positive pressure and finite temperature. However,
Eqs.~(\ref{mf}) and (\ref{coefficients}) do not help us understand
why the phase diagram has an accessible liquid-liquid critical
point only for limited ranges of $w_A$ and $U_R$, given a value of
$w_R$.

To gain some insight into this point, we observe that if we plot
Eq.(\ref{mf}) with $A=0$ and no fitting parameters
(Fig.~\ref{11}), we get a rough approximation of the calculated
$U^*$ that becomes fair for the largest $w_R$. This suggests that
as a first approximation we can assume that $A=0$ at least for
$w_R/a>1$, which is consistent with the conclusion of
Ref.~\cite{anna} that in order to have two accessible critical
points in the fluid phase, the attractive and repulsive part of
the potential must compensate, i.e., $U_Av_A\simeq U_Rv_R$ or
$A\simeq 0$.

Hence from Eqs.~(\ref{va}) and (\ref{vr}) we get the approximation
\begin{equation}
\frac{U_R}{U_A}=\frac{(a+w_R+w_A)^3-(a+w_R)^3}{(a+w_R)^3-a^3} ~.
\label{atA=0}
\end{equation}
First we observe that to get an accessible liquid-liquid critical
point, $U_R/U_A\sim O(1)$ is the relevant case. Indeed, the case
with $U_R/U_A\gg 1$ at high-enough $T$ and small-enough $P$
corresponds to an effective attractive potential with no repulsive
shoulder and a hard core at a distance of $a+w_R$ with no
liquid-liquid phase transition, or with a liquid-liquid phase
transition at vanishing $T$ and very high $P$ (see MD results in
Fig.s~9h, 9i in Ref.~\cite{anna}). On the other hand, for
$U_R/U_A\simeq 0$, Eq.~(\ref{atA=0}) gives $w_A\simeq 0$, leading
to a simple hard-core potential with no attractive well. Hence we
consider the case with $U_R/U_A\sim O(1)$.

Next, we observe that for increasing $w_R$ ($w_R\rightarrow
\infty$), Eq.~(\ref{atA=0}) goes to
\begin{equation}
\frac{U_R}{U_A}=\left(1+\frac{w_A}{a+w_R}\right)^3-1 ~,
\label{eq7}
\end{equation}
from which the condition $U_R/U_A \simeq 3 w_A/w_R$ follows for
small $w_A/w_R$. This relation is reasonably satisfied by HNC data
for $w_R/a=1$ ($w_A/a \simeq 0.4$ for $U_R/U_A \simeq 0.9$) and is
better approximated for $w_R/a=1.5$ ($w_A/a \simeq 0.6$ for
$U_R/U_A \simeq 0.9$). We can deduce from these considerations
that to get a phase diagram with two accessible critical points in
the fluid region, the three parameters of the potential should be
related by the approximate relation in Eq.~(\ref{eq7}) for $w_R/a
\gg 1$, which reduces to $w_A \simeq w_R /3$ for $U_R/U_A \simeq
1$.

\section{Conclusions}

The purpose of the present investigation is to understand the role
that the different components of the interparticle interaction
play in the physical mechanism underlying the liquid-liquid phase
transition in one-component systems. Thus we investigate the phase
diagram associated with an isotropic pair potential with an
attractive well and a repulsive shoulder, described by three
parameters, by analyzing for which combinations of the parameters
do we find a phase diagram with two critical points in the fluid
phase. In a first paper \cite{anna} we used MD simulations finding
limited ranges of the parameters so that the liquid-liquid phase
transition was accessible. We also presented a general description
based on the MVDWE approach, which rationalized our MD results.

Since MD simulations are precise but time-demanding, we completed
this analysis by adopting a different approach with well-known
limitations, but one that was extremely fast in terms of
computational time, consisting of integral equations in the HNC
approximation. It is important to stress that the drawbacks of the
HNC equation are not critical for our purposes. Indeed, we found
that the theory, though at best only in qualitative agreement with
MD simulations, correctly reproduces the trend according to which
the simulated critical points move in the $\rho,T$ plane as the
potential parameters are changed \cite{anna}. On this basis, we
use the theoretical results to estimate the phase behavior of our
system over a portion of the parameter space much wider than that
explored by numerical simulations.

Our findings, both with MD and HNC approach, show that only a
limited number of combinations of potential parameters can be
associated with a phase diagram with two accessible critical
points in the fluid phase. A general conclusion is that the
repulsive component of the potential must equilibrate the
attractive component. By comparing the MVDWE predictions with the
results presented here, we can quantify the previous general
statement with the relation $|A|\lesssim U_Av_R$ with quantities
defined in Eqs.~(\ref{eq1})--(\ref{vr}).

For $w_R\ll a$ it is difficult to extract a clear relation between
the potential's parameters. However, we note that Eq.~(\ref{mf})
shows a leading linear relation between $U_R/U_A$ and $w_A/w_R$
for $w_R\ll a$ and $w_A\ll a$, suggesting that the liquid-liquid
phase transition could also be found in systems with short
repulsive range, if the attractive range is short as well.

\begin{figure}
\centering
\includegraphics[width=11.0cm,angle=270]{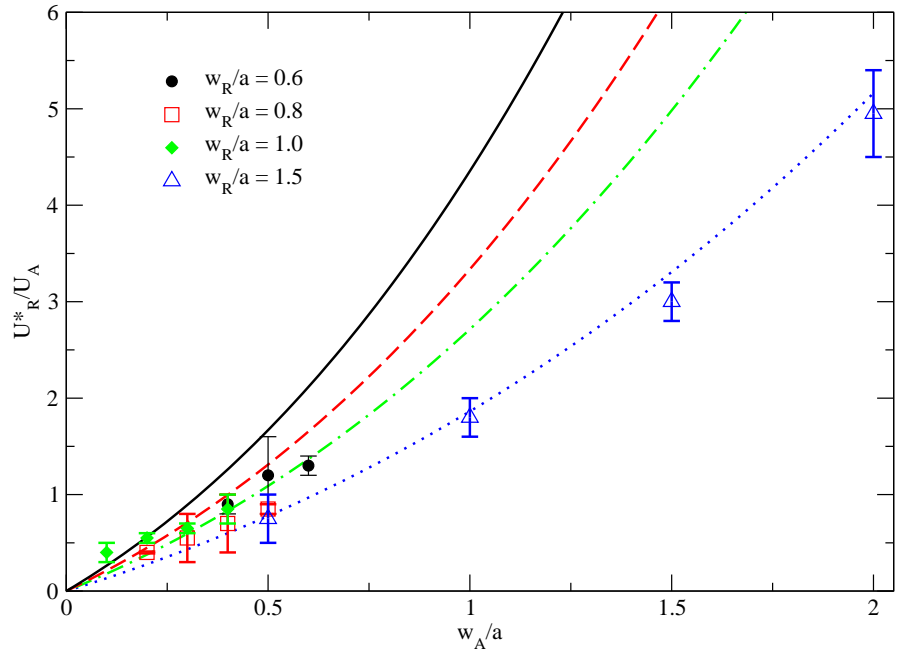}
\caption{Symbols are as in Fig.~\ref{10}. Lines are Eq.(\ref{mf})
evaluated for $A=0$: continuous line for set 1 (circles), dashed
line for set 2 (squares), dot-dashed line for set 3 (diamonds),
dotted line for set 4 (triangles).} \label{11}
\end{figure}

For $w_R\gg a$ the situation is more clear. The MVDWE predictions
for $A=0$ compare well with the HNC results (Fig.~\ref{11}),
leading to the Eq.~(\ref{atA=0}). This equation gives us the
intuitive understanding that the repulsive and attractive
components of the interaction potential compensate when the
attractive volume, weighted by the attractive energy, is equal to
the repulsive volume, weighted by the repulsive energy. Moreover,
for large $w_R$, Eq.~(\ref{atA=0}) reduces to the simple
Eq.~(\ref{eq7}), whose leading order is $3w_A/w_R$.

In conclusion, we have studied a large number of parameter
combinations of a general squared isotropic pair potential with an
attraction and soft-core repulsion by using the HNC equation. We
verify the general trend previously observed in MD simulations and
extend the analysis to a larger number of parameter combinations,
with more emphasis on cases with wide soft-core repulsion,
particularly difficult to simulate by MD. By using the MVDWE
approach introduced in Ref.~\cite{anna} to interpret the HNC
results, we find that the condition $A=0$ is well verified for
potentials with a large repulsive range and two maxima in the IL.
We expect that this condition for a liquid-liquid phase
coexistence could be generalized to a continuous isotropic
attractive potential $U(r)$ with a wide soft-core repulsion such
as
\begin{equation}
\int_a^\infty U(r) ~ d\vec{r} \simeq 0 ~, \label{integral}
\end{equation}
where $a$ is the hard-core distance. In particular, this condition
brings to Eq.~(\ref{atA=0}) two accessible critical points in the
fluid phase between the parameters of squared potentials
considered here.

{\bf Acknowledgments}

We thank P.G. Debenedetti and F. Sciortino for fruitful
discussions. G. F. acknowledges financial support from the
Ministerio de Ciecia y Tecnolog\'{\i}a (Spain) and the allocation
of computer resources from INFM Progetto Calcolo Parallelo. We
thank the NSF for financial support, through the computational
chemistry program and the collaborative research in chemistry
program.

\end{document}